\documentclass[12pt,pre,aps,showpacs,preprint]{revtex4}

\usepackage{graphics}
\usepackage{graphicx}
\usepackage{amsfonts}
\usepackage{textcomp}
\usepackage{amssymb}
\usepackage{amsmath}

\begin{document}

\title{Analytical Constructions of a Family of Dense Tetrahedron Packings and the Role of Symmetry}

\author{S. Torquato$^{1,2,3,4,5}$ and Y. Jiao$^5$}


\affiliation{$^1$Department of Chemistry, Princeton University,
Princeton New Jersey 08544, USA}




\affiliation{$^2$Princeton Center for Theoretical Physics,
Princeton University, Princeton New Jersey 08544, USA}

\affiliation{$^3$Program in Applied and Computational Mathematics,
Princeton University, Princeton New Jersey 08544, USA}

\affiliation{$^4$School of Natural Sciences, Institute for
Advanced Study, Princeton NJ 08540}

\affiliation{$^5$Department of Mechanical and Aerospace
Engineering, Princeton University, Princeton New Jersey 08544,
USA}


\begin{abstract}

The determination of the densest packings of
regular tetrahedra (one of the five Platonic solids)
is attracting great attention as evidenced by the rapid
pace at which packing records are being broken
and the fascinating packing structures that have emerged.
We have discovered the densest known packings of regular tetrahedra
with a density $\phi= \frac{12250}{14319} = 0.855506\ldots$.
These packings are special cases of an analytical
two-parameter family of dense periodic packings with
four particles per fundamental cell that we have constructed
here.
From this family, we can recover
a set of recent packing arrangements due to Kallus {\it et al.}
with density  $\phi=\frac{100}{117}=0.854700\ldots$, which
has higher symmetry than our densest packings.  We also
describe a procedure that could lead to rigorous upper bounds
on the maximal density of tetrahedron packings, which
could aid in assessing the packing efficiency of candidate dense
packings.
\end{abstract}

\pacs{61.50.Ah, 05.20.Jj}

\maketitle

\section{Introduction}

Dense packings of nonoverlapping solid objects or particles are ubiquitous
in synthetic and natural situations. Packing problems arise
in technological contexts, such as the packaging industries, agriculture (e.g., grains in silos),
and solid-rocket propellants, and underlie the structure of a multitude of biological systems (e.g., tissue
structure, cell membranes, and phyllotaxis). Dense
particle packings are intimately related to the structure
of low-temperature states of condensed matter, such
as liquids, glasses and crystals \cite{We71,Za83,Chaik95,To02a}.  In the last decade,
scientific attention has broadened from the study of dense packings of
spheres (the simplest shape that does not tile Euclidean space) \cite{Ga31,Be60,Co93,Li98,To00b,Oh02,Co03,Ha05,To06b,Co09}
to dense packings of nonspherical particles \cite{Be00,Do04d,Ji09a}.
A basic characteristic of a packing in $d$-dimensional Euclidean space $\mathbb{R}^d$
is its density $\phi$, defined to be the fraction of $\mathbb{R}^d$
that is covered by the particles.

A problem that has been of great scientific interest for
centuries is the determination of the densest arrangement(s) of
particles that do not tile space and the associated maximal
density $\phi_{max}$.
For generally shaped particles, finding the densest packings
is notoriously difficult. This salient point is summarized well
by Henry Cohn \cite{Co09b} who recently remarked, ``For most grain shapes
we cannot guess or even closely approximate
the answer, let alone prove it, and it is difficult
to develop even a qualitative understanding of
the effects of grain shape on packing density.''
Until recently, very little was known about the densest packings of
polyhedral particles. The difficulty in obtaining dense packings of
polyhedra is related to their complex rotational degrees of freedom and to the
non-smooth nature of their shapes \cite{To09b,To09c}.

Recently, we set out to attempt to determine the densest known
packings of the Platonic and Archimedean solids \cite{To09b,To09c}.
It was shown that the central symmetry of
the majority of the Platonic and Archimedean solids
distinguish their dense packing arrangements from
those of the non-centrally symmetric ones in a
fundamental way. (A particle is centrally symmetric if it has a center $C$
that bisects every chord through $C$ connecting any two
boundary points of the particle; i.e., the center is a point of
inversion symmetry.) The tetrahedron is the only Platonic solid
that lacks central symmetry, an attribute that geometrically frustrates
it to a greater degree than the majority of the remaining solids
in this set that do not tile space \cite{To09c}.
A number of organizing principles
emerged in the form of conjectures for polyhedra as well
as other nonspherical shapes. In the case of polyhedra,
the following two are particularly relevant:

\begin{itemize}
\itemsep 0.05in
\item {\bf Conjecture 1}: {\sl The densest packings of the centrally symmetric
Platonic and Archimedean solids are given by their corresponding optimal Bravais lattice packings.}

\item {\bf Conjecture 2}: {\sl The densest packing of any convex, congruent polyhedron
without central symmetry generally is not a (Bravais) lattice packing, i.e.,
set of such polyhedra whose optimal packing
is not a Bravais lattice is overwhelmingly larger than the set whose optimal
packing is a Bravais lattice.
}
\end{itemize}

Conjecture 1 is the analog of Kepler's sphere conjecture for the centrally symmetric
Platonic and Archimedean solids. In this sense, such  solids behave
similarly to spheres in that their densest packings are lattice arrangements
and (except for the cube) are geometrically frustrated like spheres.
Conjecture 2 has been shown by Conway and Torquato \cite{Co06} to be true for both the tetrahedron
and Archimedean non-centrally symmetric truncated tetrahedron, the latter
of which can be arranged in a ``uniform" non-Bravais lattice packing with density at least as high as $23/24=0.958333\ldots$. (A uniform packing, defined more precisely below,
has a  symmetry that takes one tetrahedron to another.)
Indeed, it was this investigation that has spurred the flurry
of activity in the last several years to find the densest packings of tetrahedra.
There have been many twists and unexpected turns since 2006 that have led
to the densest known packings of tetrahedra that we report here with
$\phi = \frac{12250}{14319} = 0.855506\ldots$, which is the
highest density obtained from our two-parameter family of constructions
described below. Therefore, to place our present results in their proper context,
it is instructive to review briefly the developments since
2006.

\begin{table}
\centering
\caption{A brief summary of the dense non-lattice packings
of  tetrahedra. The name of the packing is given along with the year
that it was discovered. Here  $\phi$ is the packing density and $N$ is the number
of tetrahedra per fundamental cell.}
\begin{tabular}{c@{\hspace{0.45cm}}c@{\hspace{0.45cm}}c@{\hspace{0.45cm}}c}
\hline\hline
Packing & Year & $\phi$ &  $N$  \\
\hline
Uniform I  \cite{Co06}  & 2006   & $\frac{2}{3}=0.666666\ldots$   & 2 \\
Welsh     \cite{Co06}      & 2006   & $\frac{17}{24}=0.708333\ldots$   & 34  \\
Icosahedral   \cite{Co06}  & 2006   & $0.716559\ldots$   & 20  \\
Uniform II \cite{Ka09} & 2009 & $\frac{139+40\sqrt10}{369}=0.719488\ldots$ & 2 \\
Wagon Wheels   \cite{Ch08} & 2008   & $0.778615\ldots$   & 18  \\
Improved Wagon Wheels \cite{To09b} & 2009 & $0.782021\ldots$ & 72 \\
Disordered Wagon Wheels \cite{To09c}& 2009 & $0.822637\ldots$ & 314 \\
Ring Stacks \cite{Gl09} &2009& $0.8503\ldots$ & 82 \\
Uniform III \cite{Ka09} & 2009 &$\frac{100}{117}=0.854700\ldots$ & 4\\
Dimer-Uniform& 2009 & $\frac{12250}{14319} = 0.855506\ldots$& 4\\
\hline\hline
\end{tabular}
\label{tab1}
\end{table}

First, we note that the densest Bravais-lattice packing of tetrahedra
(requiring one tetrahedron per fundamental cell such that
each tetrahedron in the packing has the same
orientation as the others) has a packing fraction $\phi =18/49=0.367\ldots$ and each tetrahedron
touches 14 others \cite{Ho70}. Conway and Torquato \cite{Co06} showed
that the densest packings of tetrahedra cannot be Bravais lattices
by analytically constructing several such packings
with densities that are substantially larger than $0.367$.
(A non-Bravais lattice packing contains
multiple particles, with generally different orientations, per fundamental cell, which is periodically replicated
in $\mathbb{R}^d$.) One such packing is a ``uniform" packing with density $\phi=2/3$
and two particles per fundamental cell. The so-called ``Welsh"
packing has a density $\phi=0.708333\ldots$ and 34 particles per fundamental
cell. Yet another non-Bravais lattice packing with density  $\phi = 0.716559\ldots$
is based on the filling of ``imaginary" icosahedra with the densest
arrangement of 20 tetrahedra
and then arranging the imaginary icosahedra in their densest lattice
packing configuration. The densities of both the Welsh and Icosahedral packings can
be further improved by certain particle displacements  \cite{Co06}.
Using imperfect ``tetrahedral" dice, Chaikin {et al.} \cite{Ch07}
experimentally generated jammed disordered packings
with $\phi \approx 0.75$.
Employing physical models and a computer algebra system, Chen \cite{Ch08}
discovered a remarkably dense periodic arrangement
of tetrahedra with $\phi=0.7786\ldots$, which exceeds
the density ($\phi_{max}=\pi/\sqrt{18} =0.7404\ldots$) of the
densest sphere packing by an appreciable amount.
We have called this the ``wagon-wheels" packing \cite{To09b,To09c}.

Torquato and Jiao \cite{To09b} devised and applied an optimization scheme,
called  the adaptive-shrinking-cell (ASC) method, that used
an initial configuration based on the wagon-wheels packing to yield
a non-Bravais lattice packing consisting of 72 tetrahedra per fundamental cell
with a density $\phi =0.782\ldots$ \cite{To09b}.
Using 314 particles per fundamental cell and starting
from an ``equilibrated" low-density liquid configuration,
the same authors were able to improve the density to $\phi =0.823\ldots$ \cite{To09c}.
This packing arrangement interestingly lacks long-range
order. Haji-Akbari {\it et al.} \cite{Gl09} numerically
constructed a periodic packing of tetrahedra made of parallel stacks of ``rings" around
``pentagonal'' dipyramids consisting of 82 particles per fundamental cell
and a density $\phi=0.8503\ldots$. More recently, Kallus {\it et al.}
\cite{Ka09} found a remarkably simple uniform packing of tetrahedra
with high symmetry consisting of only four particles per fundamental
cell with density $\phi=\frac{100}{117}=0.854700\ldots$. Table \ref{tab1} summarizes some of the
packing characteristics of the non-Bravais lattice
packings of tetrahedra.

\section{Two-Parameter Family of Dense Packings of Tetrahedra}

Inspired by the work of Kallus et al. \cite{Ka09}, we have applied
the adaptive-shrinking-cell (ASC)
optimization scheme to  examine comprehensively packings with a considerably 
small number of particles per fundamental cell (from 2 to 32) than we  have used in the past.
The ASC scheme employs
both a sequential search of the configurational space of the
particles and the space of lattices via an adaptive fundamental
cell that deforms and shrinks on average to obtain dense packings
A dense packing with 8-particle basis that emerged from this
numerical investigation suggested that it was composed of two very similar fundamental cells,
each containing 4 particles. Using one of the 4-particle basis configurations, we were able
to find packings with density $\phi = 0.8551034\ldots$ that exceeded the highest density
packings with $\phi=100/117=0.854708\ldots$ constructed by Kallus {\it et al.}
Even though our packings  possess a type of point inversion symmetry,
they are not as symmetric as the densest packings reported in Ref. \cite{Ka09}, as we now explain. 
The four tetrahedra in the fundamental cell
in our dense numerically generated packings formed two contacting ``dimers''.
A dimer is composed of a pair of regular tetrahedra with unit edge length
that exactly share a common face.
The compound object consisting of the two contacting dimers possesses point inversion symmetry,
with the inversion center at the centroid of the contacting region on the faces.
A Bravais lattice possesses point inversion symmetry about the lattice points and the
centroids of the fundamental cells. By placing the symmetry center of the
two-dimer compound on the centroids (or the lattice points),
we construct packings that generally possess point inversion symmetry
only about the symmetry centers of the two-dimer compound.
We call such structures  {\it dimer-uniform} packings,
since the inversion symmetry acts to take any {\it dimer} into another. 
Such packings should be distinguished from the more symmetric {\it uniform} (or transitive) packings of 
tetrahedra in which the symmetry operation acts to take any {\it tetrahedron} into another,
such as the ones found in Refs.~\cite{Co06} and ~\cite{Ka09} (see Table~\ref{tab1}).
The latter have almost as much  symmetry as a Bravais lattice,
except that the centroids of the particles are not just characterized by
simple translational symmetry.


We then set out to obtain analytical constructions based on our numerical packings
that relax the symmetry conditions on the contacting dimers. In particular, we orient
the 3-fold rotational symmetry axis of one of the dimers in an arbitrary direction
(say the $z$-direction of a Cartesian coordinate system), and
then fix the origin of the lattice vectors at the centroid of this dimer.
(The centroid is located at the center of the contacting faces of the two tetrahedra
that comprise the dimer.)
Then we place the second dimer in contact with the first one such that there is a
center of inversion symmetry that takes one dimer to the other, which implies
face-to-face contacts between the two dimers.

The problem of determining the analytical constructions then amounts to determining
12 equations for the 12 unknowns. Nine of the 12 unknowns arise from the three unknown lattice
vectors, each of which contains three unknown components. The other 3 unknowns
derive from the components of the centroid of the second dimer.

In particular, we let the centroid of the dimer at the origin denoted by ${\bf v}_o$,
and the centroid of the other dimer denoted by ${\bf v}_c$. The vertices of the two dimers
at ${\bf v}_o$ and ${\bf v}_c$ are given by ${\bf r}_A = (\frac{1}{2}, \frac{1}{2\sqrt3},0)$,
${\bf r}_B = (-\frac{1}{2}, \frac{1}{2\sqrt3},0)$, ${\bf r}_C = (0, -\frac{1}{\sqrt3},0)$,
${\bf r}_D = (0, 0, \sqrt{\frac{2}{3}})$, ${\bf r}_E = (0, 0, -\sqrt{\frac{2}{3}})$ and
${\bf r}^*_A = -{\bf r}_A+{\bf v}_c$, ${\bf r}^*_B = -{\bf r}_B+{\bf v}_c$, ${\bf r}^*_C = -{\bf r}_C+{\bf v}_c$,
${\bf r}^*_D = -{\bf r}_D+{\bf v}_c$, ${\bf r}^*_E = -{\bf r}_E+{\bf v}_c$, respectively.
In addition, let the lattice vectors be ${\bf v}_1$, ${\bf v}_2$, and ${\bf v}_3$.
The 12 components of the four vectors (${\bf v}_i$, i=c,1,2,3) are the aforementioned unknowns
that are related to each other through the nonoverlapping conditions.

In our packings, each
dimer has 8 face-to-face contacts and \textit{at least} 2 edge-to-edge contacts.
Among these 20 contacts of the two dimers in the fundamental cell,
there are 8 independent face-to-face contacts and 1 independent edge-to-edge contact,
which reduces the number of variables for the packing from 12 to 3.

A face-to-face contact requires that the projection of the vector distance between the
centroids of the two dimers on the contacting face normal is a constant.
The 8 independent face-to-face contacts are between the dimer pairs with the
centroids at $\{{\bf v}_c, {\bf v}_o\}$, $\{{\bf v}_c, {\bf v}_o+{\bf v}_1\}$, $\{{\bf v}_c, {\bf v}_o+{\bf v}_2\}$,
$\{{\bf v}_o+{\bf v}_3, {\bf v}_c\}$, $\{{\bf v}_o+{\bf v}_3, {\bf v}_c-{\bf v}_2\}$,
$\{{\bf v}_o+{\bf v}_3, {\bf v}_c+{\bf v}_1-{\bf v}_2$\}, $\{{\bf v}_o, {\bf v}_c-{\bf v}_1+{\bf v}_3\}$,
and $\{{\bf v}_o, {\bf v}_c+{\bf v}_1-{\bf v}_2-2{\bf v}_3\}$. The contact between
dimer pairs at $\{{\bf v}_i, {\bf v}_j\}$ requires
\begin{equation}
\label{eq001}
({\bf v}_i-{\bf v}_j)\cdot {\bf n}_{ij} = \frac{2\sqrt6}{9},
\end{equation}
where ${\bf n}_{ij}$ is unit outward contacting face normal of the dimer at ${\bf v}_j$.

The edge-to-edge contact requires that the projection of the vector connecting the corresponding ends of
two edges on the common perpendicular line of the two edges equals zero, i.e.,
\begin{equation}
\label{eq002}
[{\bf r}_A-({\bf r}_B-{\bf v}_1+{\bf v}_2+{\bf v}_3)]\cdot{\bf l}_0 =0 ,
\end{equation}
where ${\bf l}_0 = ({\bf r}_A-{\bf r}_D)\times({\bf r}_B-{\bf r}_D)$.

Moreover, there are two independent nonoverlapping conditions obtained from 4 \textit{possible}
edge-to-edge contacts between neighboring particles, i.e.,
\begin{equation}
\label{eq003}
[{\bf r}_C-({\bf r}_B-{\bf v}_1+{\bf v}_3)]\cdot{\bf l}_1 \ge 0 ,
\end{equation}
\begin{equation}
\label{eq004}
[{\bf r}_C-({\bf r}_D+{\bf v}_3)]\cdot{\bf l}_2 \ge 0 ,
\end{equation}
where ${\bf l}_1 = ({\bf r}_B-{\bf r}_D)\times({\bf r}_E-{\bf r}_C)$
and ${\bf l}_2 = ({\bf r}_D-{\bf r}_A)\times({\bf r}_C-{\bf r}_E)$.

Furthermore, there are two additional nonoverlapping conditions given by 4 potential vertex-to-face contacts, i.e.,
\begin{equation}
\label{eq005}
{\bf v}_2 \cdot {\bf n}_1 = \sqrt{\frac{2}{3}},
\end{equation}
where ${\bf n}_1 = (\frac{\sqrt6}{3}, -\frac{2\sqrt2}{6}, -\frac{1}{3})$ is unit outward normal
of the contacting face,
\begin{equation}
\label{eq006}
({\bf v}_1-{\bf v}_2) \cdot {\bf n}_2 = \sqrt{\frac{2}{3}},
\end{equation}
and ${\bf n}_2 = (-\frac{\sqrt6}{3}, -\frac{2\sqrt2}{6}, \frac{1}{3})$ is unit outward normal
of the contacting face.

The edge-to-edge and vertex-to-face contacts are realized when the
equality holds in the above conditions (\ref{eq002})-(\ref{eq006}).
However, these contacts can not be realized simultaneously in general.
In particular, Eqs.~(\ref{eq001}), (\ref{eq002}), (\ref{eq003}), (\ref{eq005}) and
Eqs.~(\ref{eq001}), (\ref{eq002}), (\ref{eq004}), (\ref{eq006}) provide two sets of equations
that lead to a family of dense tetrahedral packings whose structures are determined by two
parameters $a$ and $b$, with the corresponding density dependent only on the parameter $a$.
In other words, each density is associated with a spectrum of different packing structures.

\begin{figure}
\begin{center}
$\begin{array}{c}
\includegraphics[height=7.5cm, keepaspectratio]{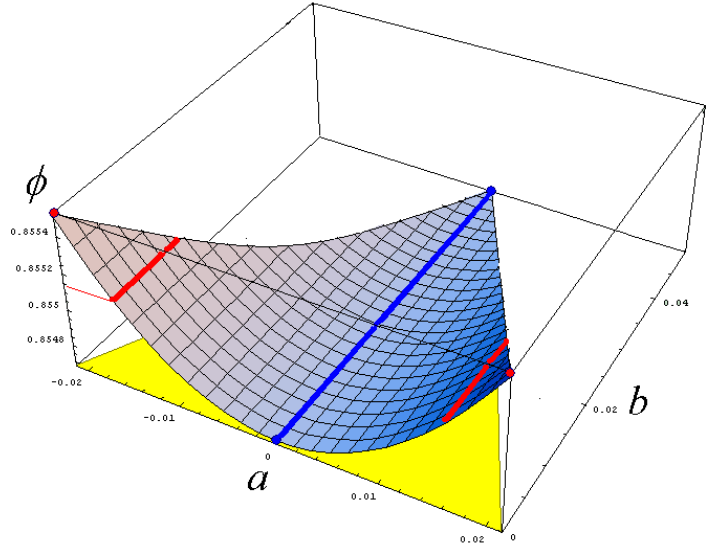} \\
\end{array}$
\end{center}
\caption{(color online). The density $\phi$ surface of our family of tetrahedral packings
as a function of the two parameters $a$ and $b$. As explained in
the text, the thick red lines show two sets of tetrahedral packings with distinct structures
but with the same density. The two red points correspond to the densest known two tetrahedral packings. The packings found by Kallus {\it et al.} \cite{Ka09} are
recovered from our two-parameter family (thick blue line).}
\label{fig1}
\end{figure}

Finally, we arrive at an explicit expression for the packing density:
\begin{equation}
\label{eq007}
\phi = \displaystyle{\frac{V_T}{|{\bf v}_1\times{\bf v}_2\cdot{\bf v}_3|}},
\end{equation}
where $V_T = \sqrt2/12$ is the volume of a regular tetrahedron with unit edge length. Substituting
the lattice vectors expressed in terms of the two parameters $a$ and $b$ into (\ref{eq007}) yields
\begin{equation}
\label{eq1}
\phi = \displaystyle{\frac{100}{117-240 a^2}},
\end{equation}
where $a \in (-\frac{3}{140},\frac{3}{140})$. It is important to note
that for each $a \neq 0$, there are two sets of
packings of tetrahedra, each with distinct structures but
possessing the same density (as shown in Fig. \ref{fig1} by the thick red lines). For
$-\frac{3}{140}<a<0$, the packings are specified by
\begin{equation}
\label{eq2}
\begin{array}{c}
\displaystyle{{\bf v}_c = (\frac{1}{5}+\frac{a}{3}, -\frac{4}{5\sqrt3}, -\frac{3\sqrt2}{5\sqrt3}+
\sqrt{\frac{2}{3}}a)},\\
\displaystyle{{\bf v}_1 = (-a, -\frac{\sqrt3}{2}, \sqrt{\frac{2}{3}}a)}, \\
\displaystyle{{\bf v}_2 = (\frac{3}{4}+a-b, -\frac{\sqrt3}{4}+\sqrt3 a, -\frac{3}{5\sqrt6}-\frac{8}{\sqrt6}a+\sqrt6 b)}, \\
\displaystyle{{\bf v}_3 = (-\frac{7}{20}-2a+b, -\frac{\sqrt3}{4}-\sqrt3 a, -\frac{9}{5\sqrt6}+\frac{10}{\sqrt6}a-\sqrt6 b)},
\end{array}
\end{equation}
where $0<b<\frac{3+140a}{60}$. For $0<a<\frac{3}{140}$, the packings are specified by
\begin{equation}
\label{eq3}
\begin{array}{c}
\displaystyle{{\bf v}_c = (\frac{1}{5}+\frac{a}{3}, -\frac{4}{5\sqrt3}, -\frac{3\sqrt2}{5\sqrt3}+\sqrt{\frac{2}{3}}a)},\\
\displaystyle{{\bf v}_1 = (-a, -\frac{\sqrt3}{2}, \sqrt{\frac{2}{3}}a)}, \\
\displaystyle{{\bf v}_2 = (\frac{3}{4}-2a-b, -\frac{\sqrt3}{4}+\sqrt3 a, -\frac{3}{5\sqrt6}+\frac{10}{\sqrt6}a+\sqrt6 b)}, \\
\displaystyle{{\bf v}_3 = (-\frac{7}{20}+a+b, -\frac{\sqrt3}{4}-\sqrt3 a, -\frac{9}{5\sqrt6}-\frac{8}{\sqrt6}a-\sqrt6 b)},
\end{array}
\end{equation}
where $0<b<\frac{3-140a}{60}$.

\begin{figure}
\begin{center}
$\begin{array}{c@{\hspace{1.5cm}}c}
\includegraphics[height=5.5cm, keepaspectratio]{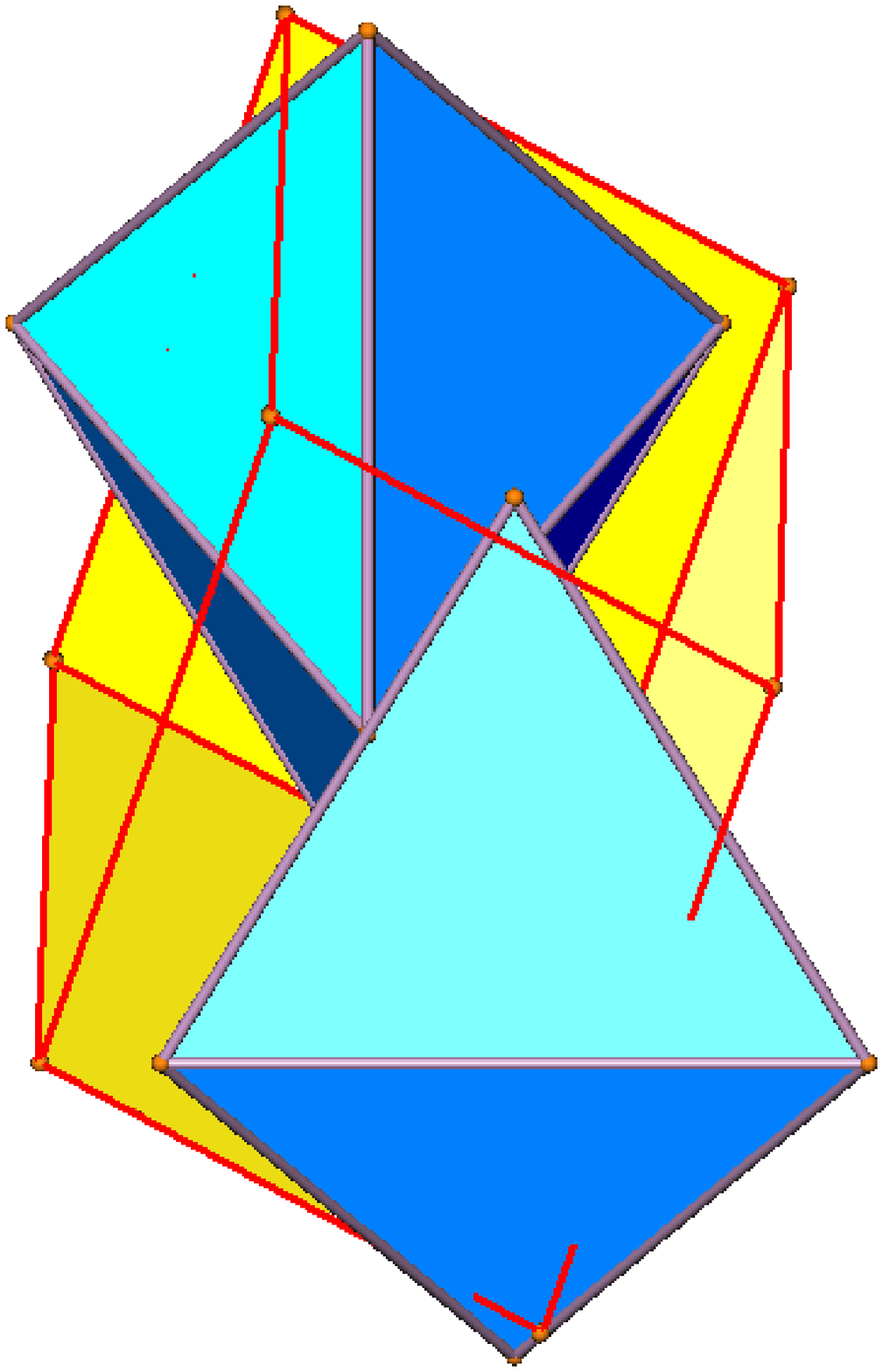} &
\includegraphics[height=5.5cm, keepaspectratio]{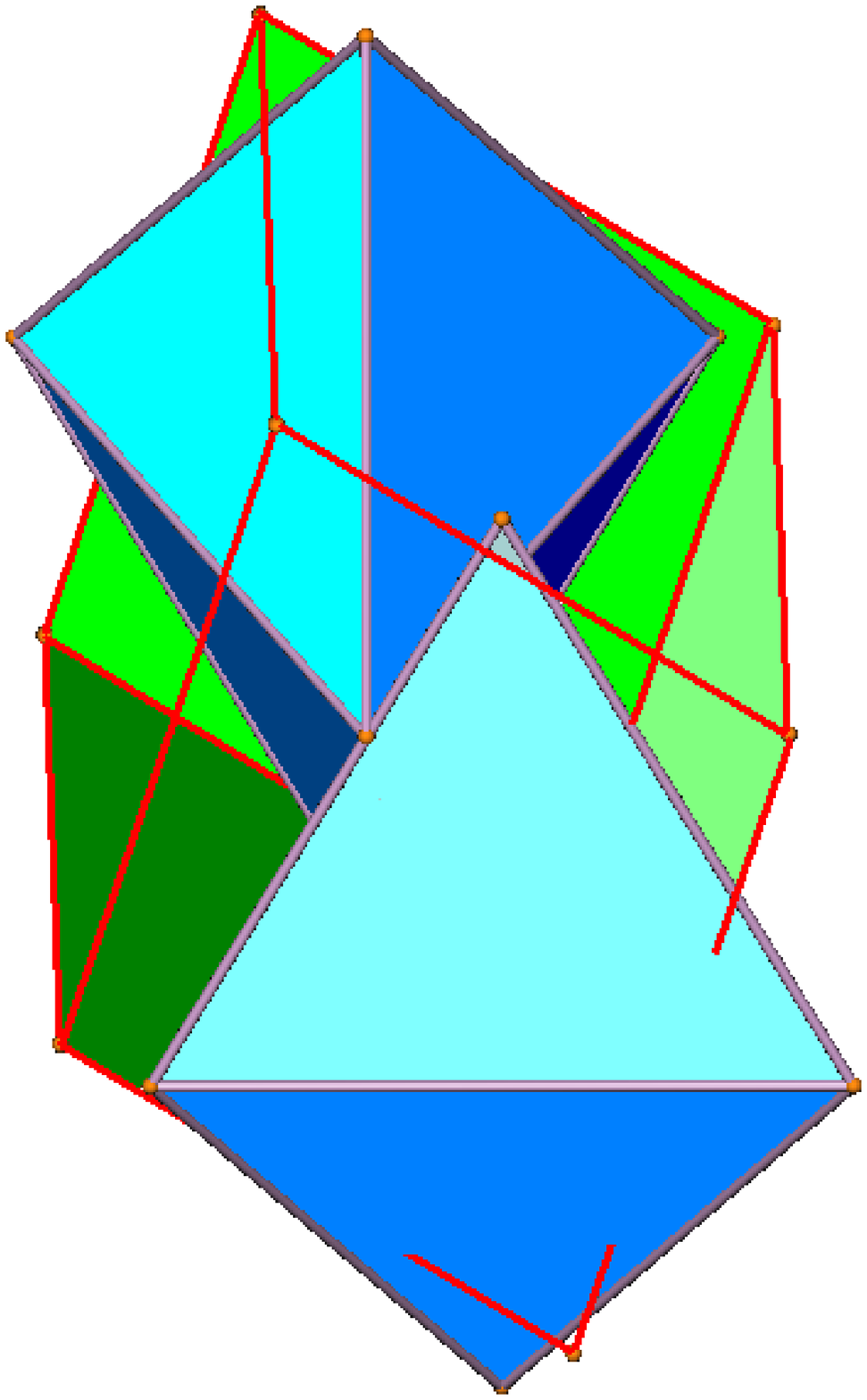} \\
\mbox{(a)} & \mbox{(b)}
\end{array}$
\end{center}
\caption{(color online). Two different configurations of the densest known packings
of  four tetrahedra (two dimers) within their corresponding rhombohedral fundamental
cells: (a) Here $a=-3/140$ and $b=0$, and the fundamental cell is colored yellow
and its boundaries are colored red. (b) Here $a=3/140$ and $b=0$, and
fundamental cell is colored blue
and its boundaries are colored red.
The two packings shown in (a) and (b) are only slightly different from one
another. Specifically,  the
difference between the coordinates of the centroids that are not at the
origin is $(1/70, 0, \sqrt6/70)$. The differences between the shapes
of the rhombohedral fundamental cells in the two cases are readily apparent.
}
\label{fig2}
\end{figure}

\begin{figure}
\begin{center}
$\begin{array}{c@{\hspace{0.25cm}}c}
\includegraphics[height=7.0cm, keepaspectratio]{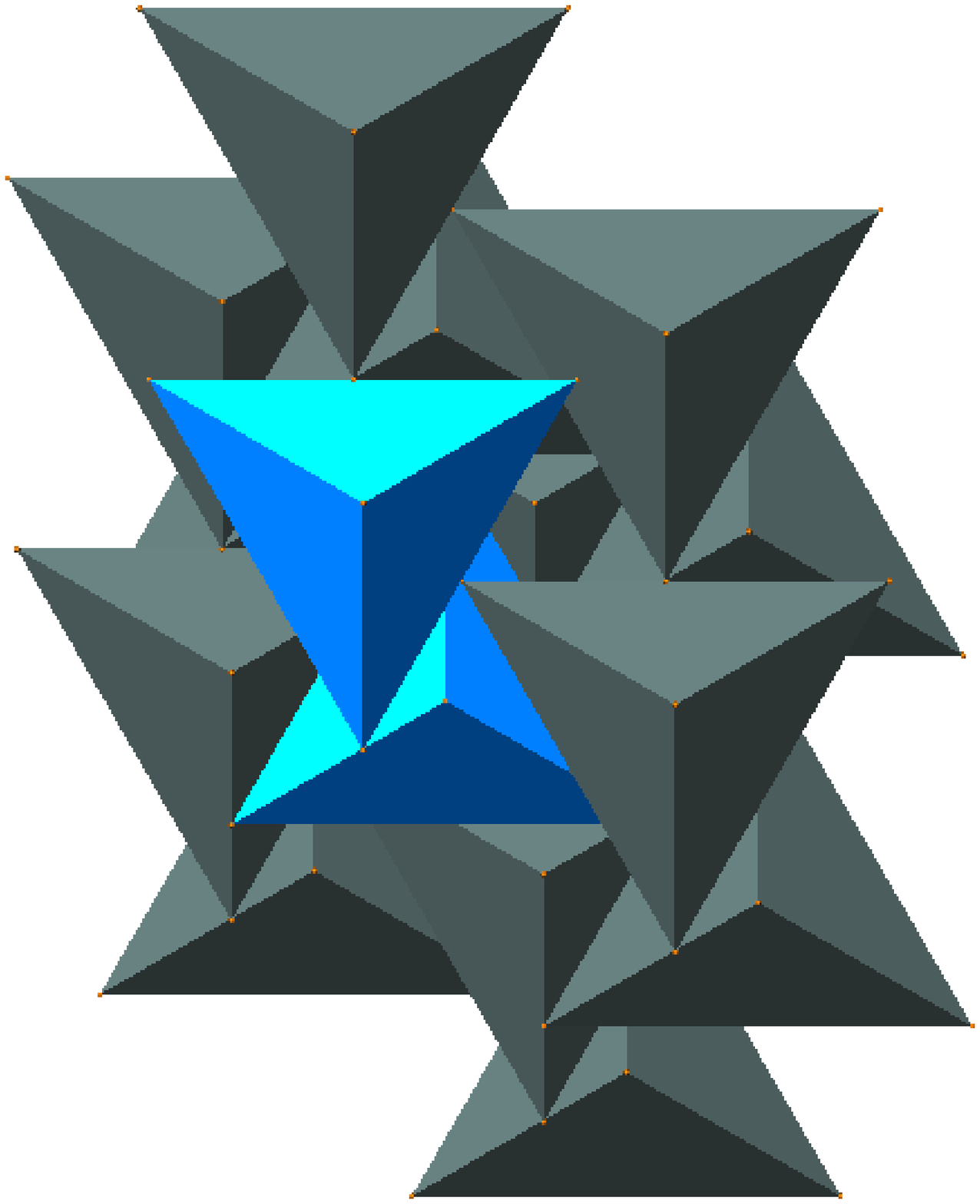} &
\includegraphics[height=7.0cm, keepaspectratio]{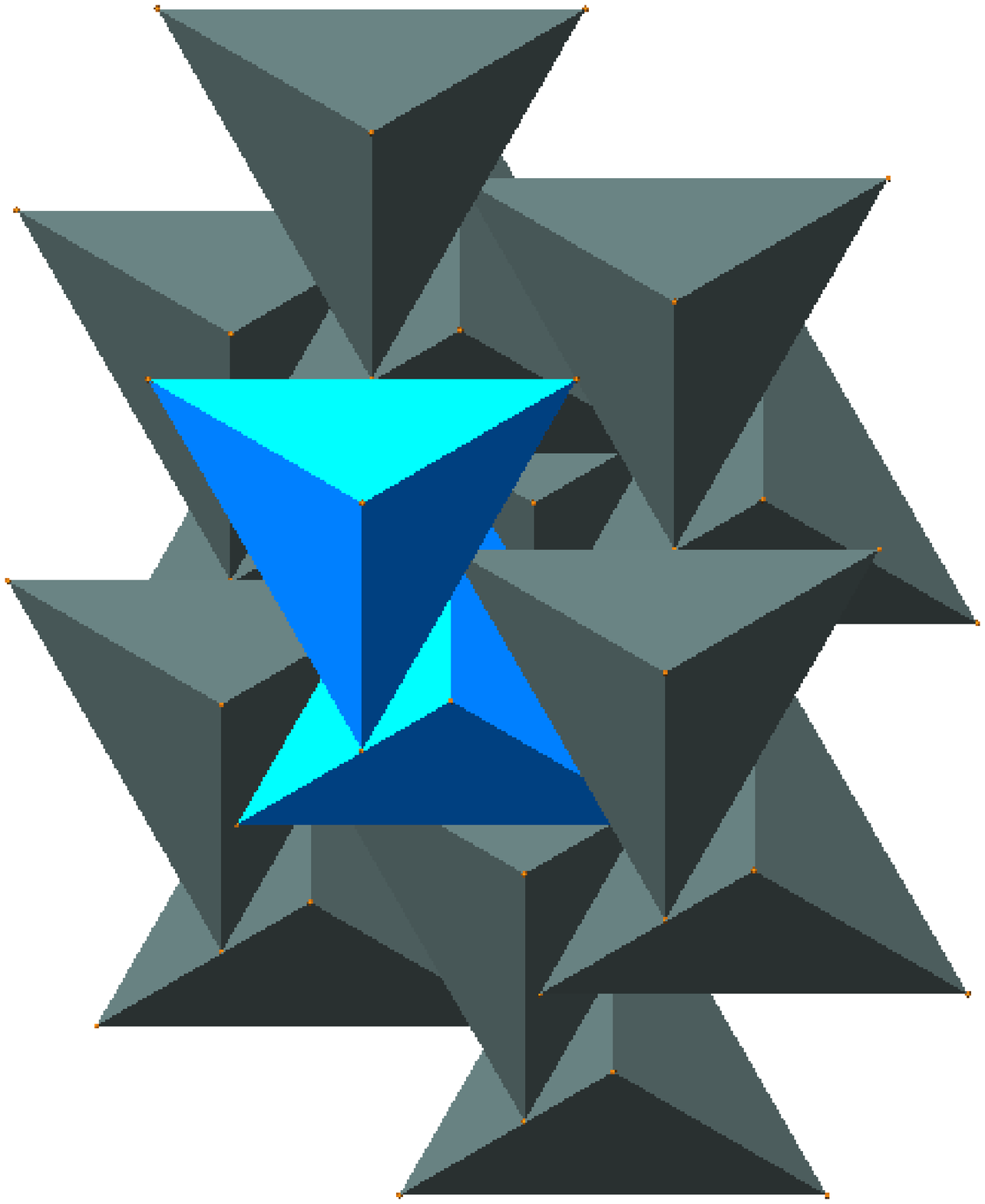} \\
\mbox{(a)} & \mbox{(b)}
\end{array}$
\end{center}
\caption{(color online). This figure shows periodic replicates the
densest known tetrahedral packings corresponding
to Fig. \ref{fig2} with 8 fundamental cells (2 along each lattice vector.
The four tetrahedra within
a fundamental cell are shown in blue: (a) $a=-3/140$ and $b=0$. (b) $a=3/140$ and $b=0$.
Note that in (a),
the dimer with centroid at ${\bf v}_1$ is slightly shifted to the
right with respect to the dimer at the origin; and in (b), the dimer
with the centroid at ${\bf v}_1$ is slightly shifted to the left with
respect to the dimer at the origin.
}
\label{fig3}
\end{figure}

The densest packings are associated with $a=-\frac{3}{140}$, $b=0$ and $a=\frac{3}{140}$, $b=0$, possessing
a density $\phi_{max} = \frac{12250}{14319} = 0.855506...$. In each set, there
is a unique packing structure associated with $\phi_{max}$ (shown as the red points
in Fig. \ref{fig1}), instead of a spectrum of structures.
 Two different configurations of the densest known packings
of  four tetrahedra within their corresponding rhombohedral fundamental
cells are shown in Fug. \ref{fig2}.
Figure \ref{fig3} depicts periodic replicates of our
optimal tetrahedral packings corresponding to those shown in
Fig. \ref{fig2} with 8 fundamental cells (2 along each lattice vector).

At $a=0$, there is only one set of packings with the same $\phi = 100/117 = 0.85470...$, whose structures are
dependent on $b$ (shown as the blue line in Fig. \ref{fig1}).
These packings reduce exactly to those discovered by Kallus et al., which
possess relatively high symmetry. In particular, $a=0$ allows the centroids of the dimers, which are related to each other
by an integer multiple of ${\bf v}_1$, to be perfectly aligned
on two of the mirror image planes of the dimers simultaneously, which leads to additional
two-fold rotational symmetry of the packing.
This additional rotational symmetry, 
together with the point inversion symmetry, leads to uniform packings with respect
to each tetrahedron (not just each dimer), i.e., the symmetry operation acts to take each 
tetrahedron into another.

\section{Towards Upper Bounds on the Maximal Density}

The problem of determining upper bounds on the maximal density
of packings of nonspherical particles is highly nontrivial,
and yet such estimates would be indispensable in assessing
the packing efficiency of a candidate dense packing, especially
if tight upper bounds could be constructed.
It has recently been shown  that
$\phi_{max}$ of a packing of congruent nonspherical particles of
volume $v_{P}$ in $\mathbb{R}^3$ is bounded from above according to
\begin{equation}
\phi_{max}\le  \mbox{min}\left[\frac{v_{P}}{v_{S}}\;
\frac{\pi}{\sqrt{18}},1\right], \label{bound}
\end{equation}
where $v_{S}$ is the volume of the largest sphere that can be inscribed
in the nonspherical particle and $\pi/\sqrt{18}$ is the maximal sphere-packing density
\cite{To09b,To09c}.
The upper bound (\ref{bound}) will be relatively tight for packings of nonspherical
particles provided that the {\it asphericity} $\gamma$
(equal to the ratio of the circumradius to the inradius)  of the
particle is not large. However, for tetrahedra, the asphericity is too large
for the upper bound (\ref{bound}) to yield a result that is less than unity.

One possible approach to obtaining nontrivial upper bounds is to attempt
to generalize the idea that Rogers used to prove upper bounds
on $\phi_{max}$ for sphere packings \cite{Ro58}. The key concept is to
consider a locally dense cluster of 4 contacting spheres
in a tetrahedral arrangement and then prove that the fraction
of space covered by the spheres within the tetrahedron
joining the sphere centers is an upper bound on $\phi_{max}$.
This can be done because one can triangulate any sphere packing to
decompose it into generally irregular tetrahedra with vertices at sphere centers.
The fact that the regular tetrahedron has the best density for any tetrahedron,
then yields an upper bound for the density of any sphere packing.
In the case of the non-tiling Platonic and Archimedean solids,
a natural choice for the enclosing region  associated
with the cluster is its convex hull.

\begin{figure}
\begin{center}
$\begin{array}{c}
\includegraphics[height=6.0cm, keepaspectratio]{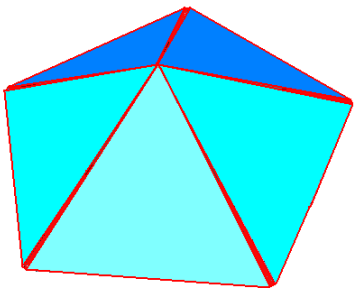} \\
\end{array}$
\end{center}
\caption{(color online). The convex hull of five regular tetrahedra in a ``wagon wheel'' arrangement.
The convex hull can be decomposed into five regular tetrahedra (shown in blue) and five
thin irregular tetrahedra (shown in red).}
\label{wagon}
\end{figure}

For tetrahedra, we must identify the {\it least} densest local cluster
with density that exceeds $\phi_{max}$. A trivial choice is a
dimer because the fraction of space covered by the dimer
within its convex hull is unity. A nontrivial choice
is a 5-particle ``wagon-wheel" cluster as shown in
Fig. \ref{wagon}.
The convex hull of the local packing of five tetrahedra sharing a common edge (a ``wagon wheel'' cluster)
can be decomposed into five regular tetrahedra and five thin irregular tetrahedra.
We assume the gaps between the regular tetrahedra are equal,
i.e, the thin irregular tetrahedra are congruent.
Note the regular tetrahedron shares two faces with its neighboring irregular tetrahedra.
Thus, the volume ratio is equal to the ratio of the corresponding heights of the regular
and irregular tetrahedron, i.e.,
\begin{equation}
\displaystyle{\gamma = \frac{V_T}{V_{T_*}} = \frac{1}{\frac{\sqrt2}{2}(3\cos^2\frac{3\pi}{10}-1)},}
\end{equation}
where $V_T$ and $V_{T_*}$ is the volume of the regular and irregular
tetrahedron, respectively. Thus,
the density of this local packing, defined as the fraction space
covered by the regular tetrahedra within the convex hull is given by
\begin{equation}
\displaystyle{\phi_W = \frac{V_T}{V_T+V_{T_*}}=\frac{1}{\frac{3\sqrt2}{2}\cos^2\frac{3 \pi}{10}+(1-\frac{\sqrt2}{2})}=0.974857\ldots.}
\end{equation}
Because this cluster is highly anisotropic and is an effectively ``flat" object,
it is reasonable to assume that it is not the least densest local
cluster and therefore its local density is a gross overestimate of $\phi_{max}$.

\begin{figure}
\begin{center}
$\begin{array}{c}
\includegraphics[height=5.5cm, keepaspectratio]{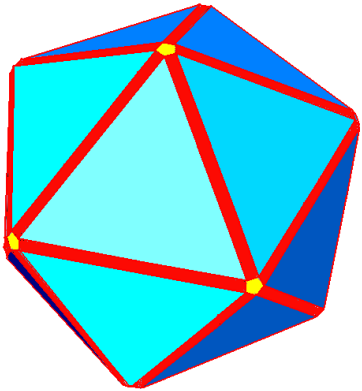} \\
\end{array}$
\end{center}
\caption{(color online). The convex hull of 20 regular tetrahedra in an ``icosahedral" arrangement.
The convex hull can be decomposed into 20 regular tetrahedra (shown in blue), 12 pyramids with
pentagonal bases (shown in yellow) and 30 pyramid with rectangular bases (shown in red).}
\label{icosa}
\end{figure}

Since tetrahedra are fully three-dimensional objects, it seems
reasonable to assume that the least densest local
cluster should be more ``isotropic" than the highly anisotropic and effectively
``two-dimensional" wagon-wheel cluster.
One plausible choice for such a cluster consists of  20 tetrahedra sharing a common vertex
(i.e., an icosahedral-like cluster).
The convex hull of this  cluster
can be decomposed into 20 regular tetrahedra, 12 pyramids with pentagonal bases and 30
pyramid with rectangular bases (see Fig. \ref{icosa}).
 We assume the gaps between tetrahedra are equal and thus the
two types of pyramids are congruent. The volume of the regular tetrahedron is $V_T = \frac{\sqrt2}{12}$ and
the volume $V_P$ of the pyramid with pentagonal base is given by

\begin{equation}
\displaystyle{V_P = \frac{5}{12}\tan(\frac{3\pi}{10}) L^2 \sqrt{1-\frac{L^2}{4 \cos^2\frac{3\pi}{10}}},}
\end{equation}
and the volume $V_R$ of the pyramid with rectangular base is given by

\begin{equation}
\displaystyle{V_R = \frac{\sqrt2}{6}L\sqrt{1-L^2}},
\end{equation}
where

\begin{equation}
\displaystyle{L=(\frac{2\sqrt2}{\tau^2}-1)\sqrt{\frac{1}{6}+\frac{\sqrt5}{18}}},
\end{equation}
and $\tau = (1+\sqrt5)/2$ is the golden ratio. Thus, the local packing density is given by

\begin{equation}
\displaystyle{\phi_I = \frac{20V_T}{20V_T+12V_P+30V_R} = 0.880755\ldots}.
\end{equation}

Note in the above calculations, we have assumed that the gaps between
the tetrahedra are equal, which is sufficient to provide  an estimate of the fraction
of space covered by the cluster within its convex hull. However,
the idea of Rogers to prove an upper bound
for sphere packings cannot be used here because there is no analogous decomposition of space
into irregular convex hulls shown in Figs. \ref{wagon} or \ref{icosa}.
A completely new idea is needed to prove that the aforementioned estimates
are bounds, but it seems plausible that they are correct bounds.
If one could prove that these clusters are indeed locally denser
than the globally densest packings, then the aforementioned
estimates provide upper bounds on $\phi_{max}$, but they
cannot be sharp, i.e., they are not achievable by any  packings.
It is noteworthy the density $\phi =0.855506\ldots$
of our our densest packings is relatively close to this
 putative upper bound density estimate
of $0.880755\ldots$.

\section{Discussion}


For all of the small systems (including 2 to 32 particles) that we investigated
using our numerical ASC scheme, the
densest packings that emerged had a 4-particle basis. Our analytical constructions
indicate that for such small systems, the highest density packings we found could
be optimal. However, there is no reason to believe that denser packings
could not be discovered by carrying out exhaustive searches
to determine the globally maximal densities of packings with successively larger
numbers of particles per fundamental cell.
Previous numerical studies have indicated that dense packings may have a large
number of particles per fundamental cell arranged in a complex fashion, e.g.,
the ``disordered
wagon-wheels" packing with $\phi = 0.822637\ldots$ \cite{To09c} and the ``ring stacks"
packing  with $\phi = 0.8503\ldots$ \cite{Gl09}. However, it is plausible
that such packings are in fact only locally optimal solutions and hence
the numerical techniques used to obtain them are incapable of extricating
themselves from these ``trapped" regions of configuration
space to find denser and more ordered structures
due to the intrinsic geometrical frustration of the tetrahedron.
This may also call into question claims made by  Haji-Akbari {\it et al.} \cite{Gl09} that their
packings, which are characterized by an effective ``quasicrystal-like" plane,
are true thermodynamic equilibrium phases of tetrahedra, especially at
high densities. Instead, our highest-density constructions suggest that uniform periodic
packings with a 4-particle basis (or even some yet unknown
denser periodic packing) and unjammmed, lower-density countyerparts
could be the stable phases at such high densities.
If the latter is correct and the putative ``quasicrystal-like" phase
truly exists at intermediate densities, then it is hard to imagine how
a quasicrystal-like structure of  tetrahedra under quasi-static compression (densification)
could rearrange to a structure with a more ordered periodic arrangement with higher symmetry.
However, it is difficult to draw such definitive conclusions
without further study.

Although there could still be tetrahedral packings denser than our constructions,
it appears that all of the evidence thus far points to the fact that the
densest tetrahedral packings cannot possess very high symmetry \cite{Co06,Ch08,To09b,To09c,Gl09} due to the lack
of central symmetry of a  tetrahedron and because tetrahedra cannot tile space \cite{To09c}.
Indeed, our densest uniform 4-particle-basis packings found in the present paper improved
upon the best  analogous packings of Kallus {\it et al.} \cite{Ka09} by relaxing
the two-fold rotational symmetry constraints they imposed.

\begin{acknowledgments}
We are grateful to Henry Cohn, Yoav Kallus and John Conway for helpful discussions.
S. T. thanks the Institute for Advanced Study for
its hospitality during his stay there.
This work was supported by the Division of Mathematical Sciences
at the National Science Foundation under Award Number DMS-0804431
and by the MRSEC Program of the
National Science Foundation under Award Number DMR-0820341.
\end{acknowledgments}



\end{document}